\documentclass[12pt,a4paper]{article}

\usepackage{epsfig}
\usepackage{amsmath,amsfonts,amssymb}
\usepackage{t1enc}
\usepackage{cite}
\usepackage{color}
\providecommand{\openone}{\leavevmode\hbox{\small1\kern-3.8pt\normalsize1}}

\newcommand{\Esix}{\text{E}_6}
\newcommand{\fbin}{fb$^{-1}$}
\newcommand{\Zlp}{Z'_\lambda}
\newcommand{\pttt}{p_T^{t \bar t}}

\newcommand{\zp}{{Z'_\lambda}}

\newcommand{\ptmiss}{p_T\!\!\!\!\!\!\!\!\not\,\,\,\,\,\,\,}

\def\ALPGEN{{\small ALPGEN}}
\def\ALPGENs{{\small ALPGEN}\ }

\newcommand{\qu}{u}
\newcommand{\qd}{d}
\newcommand{\qubar}{\bar u}
\newcommand{\qdbar}{\bar d}
\newcommand{\fbar}{\bar f}

\begin{document}

\begin{center}
\begin{Large}
{\bf Combined analysis of $Z' \to t \bar t$ and $Z' \to t \bar t j$ production \\[2mm]
for vector resonance searches at LHC}
\end{Large}

\vspace{0.5cm}
F. del Aguila$^a$, J. A. Aguilar--Saavedra$^a$, M. Moretti$^b$, F. Piccinini$^c$, 
R. Pittau$^a$, \\ M. Treccani$^a$  \\[0.2cm] 
{\it $^a$ Departamento de F\'{\i}sica Te\'orica y del Cosmos and CAFPE, \\
Universidad de Granada, E-18071 Granada, Spain} \\[0.1cm]
%
{\it $^b$ Dipartimento di Fisica, Universit\`{a} di 
Ferrara, and INFN, 44100 Ferrara, Italy} \\[0.1cm]
{\it $^c$ INFN, Sezione di Pavia, v. A. Bassi 6, I~27100, Pavia, Italy
} \\[0.1cm]
\end{center}

\begin{abstract}
We have implemented a code for $Z'$ + $n$ jets production in ALPGEN, 
with $Z'$ decays into several final states, including $\ell^+ \ell^-$ and $t \bar t$. The MLM prescription is used for matching the matrix element
with the parton shower, including in this way the leading soft and collinear
corrections. In order to demonstrate its capabilities, we perform a combined analysis 
of $Z' \to t \bar t$ and $Z' \to t \bar t j$ production for a heavy 
leptophobic gauge boson. It is found that the effect of the extra jet cannot only be accounted for by a $K$ factor multiplying the leading-order cross section. In fact, the combined analysis for $Z' \to t \bar t$ and $Z' \to t \bar t j$ presented improves the statistical significance of the 
signal by 25\% ($8.55\sigma$ versus $6.77\sigma$ for a $Z'$ mass of 1 TeV), compared with the results of an inclusive analysis carried out on the same sample of $t \bar t+t \bar t j$ events. 
\end{abstract}

\section{Introduction}

Searching for neutral vector resonances is one important task in the Large Hadron Collider
(LHC) programme, being the Drell-Yan process 
$p p\rightarrow Z' \rightarrow \ell^+\ell^-\ , \ell = e, \mu ,$ 
the preferred one in these studies \cite{Langacker:2008yv,Aad:2009wy,
Ball:2007zza,Buttar:2008jx}. 
This signal with charged leptons in the final state has smaller 
backgrounds than those with only hadrons, but it requires 
that the new boson couples to both quarks and leptons, 
being this process suppressed when either of these types of 
couplings vanishes. The $t \bar t$ channel
is an alternative if 
the couplings to leptons are the ones which are negligible. 
Among the hadronic final states, $t\bar t$ is very interesting by itself 
because, being the top very heavy, it allows for a relatively easy 
identification and reconstruction, and for this reason its backgrounds
are relatively smaller. Even more, $Z' \to t \bar t$ is not only an 
alternative to the Drell-Yan process when the new 
resonance is leptophobic, but it is always complementary to 
determine the model because it involves a different combination 
of couplings, and it also allows for asymmetry measurements 
in the semileptonic decay \cite{:2007qb}. 

New gauge bosons are predicted in many of the best motivated 
Standard Model (SM) extensions. For instance, parity restoration, 
which can be at the TeV scale,  
requires extending the SM gauge symmetry to 
$\text{SU}(2)_L \times \text{SU}(2)_R \times \text{U}(1)_{B-L}$, with new neutral 
and charged gauge bosons at this scale 
\cite{Langacker:1984dc}. 
Grand unified models always predict new gauge bosons, 
and some of them may survive at lower energies 
\cite{Hewett:1988xc}. 
In general, models addressing the hierarchy problem 
which are not based on supersymmetry, such as Little Higgs models 
\cite{Diener:2009vq} or 
models with large extra dimensions
\cite{Antoniadis:1994yi}, also
introduce (a plethora of) new vector bosons, with some of them 
at the LHC reach.   
Many of them couple to quarks and leptons but, as already stressed, 
even in this case (and obviously in the leptophobic limit)  
it is important to perform a detailed resonance search 
in the $t\bar t$ channel.  

With this purpose in mind, we have extended \ALPGEN~\cite{Mangano:2002ea} 
with a generator for $Z'$ production, 
including real emissions matched with leading logarithmic (LL) 
corrections (both real and virtual ones).
This is presented in the next section, and applied to the search 
of vector resonances decaying into $t\bar t$ in the 
following ones. For the sake of illustration, we concentrate on the 
case of a leptophobic $Z'_\lambda$, which also decays into new heavy 
neutrinos $Z'_\lambda \rightarrow NN$ if $m_N < M_{Z'_\lambda}/2$ 
\cite{delAguila:2007ua}. 
The model is described in section~\ref{model}, where we also give details of the simulation of 
top pair production mediated by neutral gauge bosons. 
Then, in section~\ref{analysis} we show the relevance of using a generator 
including these contributions: a combined analysis 
of $Z'_\lambda \to t \bar t$ and $Z'_\lambda \to t \bar tj$ improves 
the LHC sensitivity to neutral vector resonances, raising the statistical 
significance of the signal by 25\% compared with an inclusive 
analysis.
Precise simulations of $Z'$ production in different channels 
are not only essential for discovery, but to determinine the 
model by measuring all the $Z'$ couplings with a precision as high 
as possible. 
The last section is devoted to our conclusions.

\section{\label{generator} The $Z'$ generator}

In this section we briefly describe the main features of the new code. 
It evaluates the matrix elements for $Z'+ n$~jets production, with $Z'$
decaying into several final states including $\ell^+ \ell^-$ and $t \bar t$. 
Actually ``$Z'+ n$~jets'' stands for the sum of the three possible intermediate 
vector bosons, namely $Z$, $\gamma^{*}$ and $Z'$;
their interferences as well as their widths and all spin correlations 
are taken into account.\footnote{The light jet(s) can also result from 
radiation off the $Z'$ decay products, as for example in 
$q \bar q \to Z' \to t \bar t g$, but for simplicity we still denote 
these processes as $Z' + n$~jets. Obviously, by $Z' \to t \bar t j$ 
we only refer to events in which the extra jet is radiated off the $Z'$ 
decay products. The radiation from top decay products is not included 
at the matrix element level.} 
Detailed information can be found in the README file available
at {\tt http://mlm.home.cern.ch/mlm/alpgen/}. 
The SM parameters are controlled, as it happens for all
the other processes implemented in \ALPGEN~\cite{Mangano:2002ea}, 
by the variable {\tt iewopt}, whose default value is {\tt iewopt = 3}. 
The possible final state is selected with the parameter {\tt ifs} 
(e.g. {\tt ifs = 0} $\to$  $e^+ e^-$, {\tt ifs = 2} $\to$ $t \bar t$), 
and its left-handed (LH) and right-handed (RH) couplings to $Z'$ 
with {\tt glzpe}, {\tt grzpe} and {\tt glzptop}, {\tt grzptop} respectively.
In addition, the new gauge boson (LH and RH) couplings to the initial quarks, 
and its mass and width can be arbitrarily defined through the variables 
{\tt glzpup, grzpup, glzpdown, grzpdown, glzpc, grzpc, glzps, grzps}, and 
{\tt masszp} and {\tt zpwid}, respectively, the lagrangian being
\begin{equation}
\mathcal{L} = -\frac{1}{2} \, \text{\tt masszp} \; (Z')^2
+ \frac{1}{2} \, \text{\tt glzpup} \; \bar u \gamma_\mu (1-\gamma_5) u \, Z'_\mu
+ \frac{1}{2} \, \text{\tt grzpup} \; \bar u \gamma_\mu (1+\gamma_5) u \, Z'_\mu + \dots \,,
\end{equation}
and {\tt zpwid} the width of the $Z'$ boson.
Finally, the MLM matching prescription~\cite{Mangano:2004,Mangano:2006rw} 
can be applied by taking the parameter {\tt ickkw = 1}.
In Table \ref{tab:zpjets_processes} we gather the parton subprocesses
evaluated for any given jet multiplicity.
\begin{table}[h]
\begin{center}
\vskip .3cm
\begin{tabular}{|c|c|c|c|}
\hline
\multicolumn{4}{|c|}{\tt Subprocesses} \\
\hline
$\qu \qubar\to f \fbar$ &$\qd \qdbar\to f \fbar$  &
$ g \qu \to f \fbar \qu  $ &$g \qd \to f \fbar \qd $\\ 
\hline
$\qu g  \to f \fbar \qu$ & $\qd g  \to f \fbar \qd$  &
$g g  \to f \fbar \qu \qubar $ &$g g  \to f \fbar \qd \qdbar $ \\
\hline
\end{tabular}
\caption{\label{tab:zpjets_processes}
Parton subprocesses (plus their charge conjugate) taken into account 
in the computation. 
Quarks $u$ and $d$ represent generic quarks of type up and down, 
respectively, and $f \fbar$ stands for the fermion pair selected 
in the $Z'$ decay through the variable {\tt ifs}. 
If the jet multiplicity exceeds the number of light quarks in the 
final state, final state gluons are added up to reach 
the desired multiplicity.}
\end{center}
\end{table}
Since our aim is studying QCD corrections to the production of 
a single $s$-channel $Z'$ boson, rather than
the production of two $Z'$ bosons, we neglect contributions with 
two light quark pairs, because in our approach the former excludes the latter.

\begin{figure}[htb]
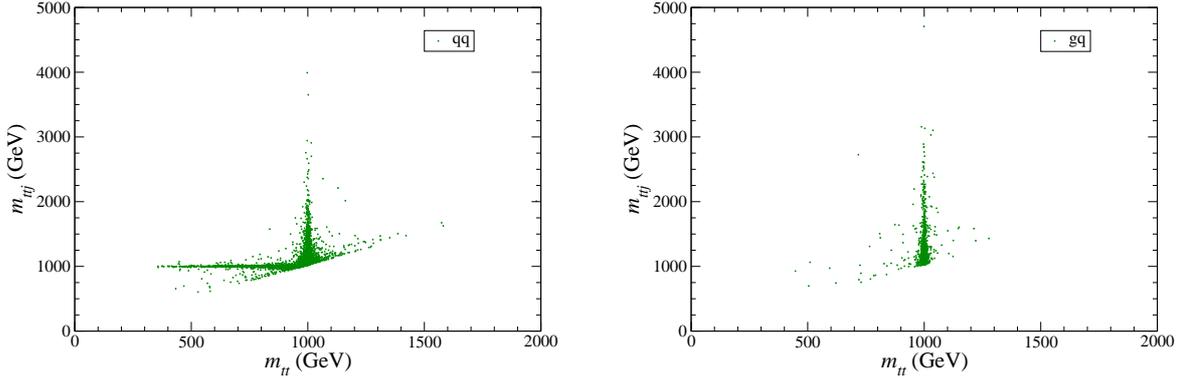

\begin{center}
\begin{tabular}{ccc}
\epsfig{file=Figs/minv-qq.eps,height=5.0cm,clip=} & \quad &
\epsfig{file=Figs/minv-gq.eps,height=5.0cm,clip=}
\end{tabular}
\caption{$t \bar t$ versus $t \bar t j$ invariant mass distributions 
at the generator level, for $Z'_\lambda j$ events originated in $q \bar q$ (left) and $gq$ (right) collisions at LHC. The number of points in the plots (8172 for $q \bar q$ and 1332 for $gq$) corresponds to the expected number of events at the LHC for $M_{Z'_\lambda}= 1$ TeV with a luminosity of 30 fb$^{-1}$.}
\label{fig:minv}
\end{center}
\end{figure}

In order to illustrate the capabilities of this code we will evaluate 
top pair production for the leptophobic model described in the 
next section. The full analysis is presented in section \ref{analysis}. 
As it can readily be seen from Fig. \ref{fig:minv},  taking $M_{Z'_\lambda} = 1$ TeV
with a total width $\Gamma_{Z'_\lambda} = 6.9$ GeV,
a proper calculation of the real radiation 
contributions to $Z'j$ is compulsory to account for the relatively large 
number of $Z'\rightarrow t\bar t j$ events produced in $q \bar q$ collisions.\footnote{Notice that in the SM, when usual isolation and transverse momentum cuts are applied, the largest contribution to $Zj$ comes from $gq$ collisions 
\cite{delAguila:2005cn}.}
(In both cases the SM contributions from $Z$ and $\gamma^*$ are 
turned off, but they are included in the analyses below.)
In particular, as it can clearly be seen in the left plot, 37\% of the $q \bar q \to t \bar t j$ events correspond to a $t \bar t j$ resonance, being the extra jet from final state radiation (FSR) from one top quark. The rest corresponds to a $t \bar t$ resonance with the jet from initial state radiation (ISR).

\section{\label{model} Description of the model and the simulation}

As explained in the previous section, the code allows for 
arbitrary $Z'$ couplings to fermions (models). 
For illustration we choose a model with vanishing couplings 
to ordinary leptons, but non-zero couplings to SM quarks 
(in particular to the top) and to new heavy neutrinos. 
Being the new boson leptophobic, the lower bound on its mass is rather weak, and
sizeable signals are already possible at Tevatron \cite{delAguila:2007ua}. 
There are many models with extra leptophobic gauge bosons, 
originally studied to interpret the initial disagreement between 
the LEP data on $Z\rightarrow b\bar b, c\bar c$~\cite{LEPrbrc} and 
the SM predictions 
\cite{Chiappetta:1996km,Babu:1996vt}. 
We will restrict ourselves to an $\Esix$ based model \cite{Gursey:1975ki}. 
The neutral gauge interactions are
described by the Lagrangian \cite{DelAguila:1995fa}
\begin{equation}
\mathcal{L}_\mathrm{NC} = - \bar \psi \gamma_\mu \left[ T_{3} g W_3^\mu
+ \sqrt{\textstyle \frac{5}{3}} Y g_Y B^\mu + Q' g' Z_\lambda^{'\mu} \right]
Â \psi \,,
\end{equation}
where a sum over the Weyl fermions of the fundamental $\Esix$ representation 
$\mathbf{27}$ and the three families must be understood.
$Y$ is the SM hypercharge properly normalised, and the
extra $\zp$ charges $Q'$ 
correspond to the only leptophobic combination within
$\Esix$ \cite{delAguila:1986iw,Babu:1996vt}, 
$ Q' = 3/\sqrt{10} \; (Y_\eta + Y/3) \,, $
with $Y_\eta$ the extra $\mathrm{U}(1)$ defined by
flux breaking \cite{Hosotani:1983xw}.
For the SM (LH) fields 
\begin{align}
& 2 Q'_u = 2 Q'_{d} = Q'_{u^c} = -2 Q'_{d^c} =
Â - \frac{1}{\sqrt{6}} \,, \nonumber \\
& Q'_{\nu} = Q'_{e} = Q'_{e^c} = 0 \, Â \, ,
\end{align}
reading the code charges for {\tt ilep} = 2 
\begin{align}
&{\tt glzptop} = {\tt glzpup} = {\tt glzpc} =
- \frac{g'}{2\sqrt{6}} \,, \nonumber \\
&{\tt grzptop} = {\tt grzpup} = {\tt grzpc} =
\frac{g'}{\sqrt{6}} \,, \nonumber \\
&{\tt glzpdown} = {\tt glzps} = {\tt grzpdown} = {\tt grzps} =
- \frac{g'}{2\sqrt{6}} \,.
\end{align}
A detailed discussion of the phenomenology of this SM
extension can be found in Ref.~\cite{Babu:1996vt}, where a nearly-leptophobic
model with $Q' \sim Y_\eta + 0.29 \, Y$ is studied together with several other
alternatives.
In our phenomenological study
we will assume for simplicity that 
the extra vector-like lepton doublets and quark singlets of
charge $-1/3$ are heavier than $M_\zp/2$, as they are the heavy
neutrinos.\footnote{If $\zp$ can decay into heavy neutrino pairs $NN$ (with 
$\zp$ charge $Q'_{N} = - {3}/{2 \sqrt{6}}$), this could be 
also observed in multi-lepton channels \cite{delAguila:2007ua,AguilarSaavedra:2009ik}.} 
Possible supersymmetric partners are taken to be heavier as well.
Otherwise, the total $Z'_\lambda$ width
would be larger, decreasing the cross sections into SM final states.
\begin{figure}[htb]
\begin{center}
\epsfig{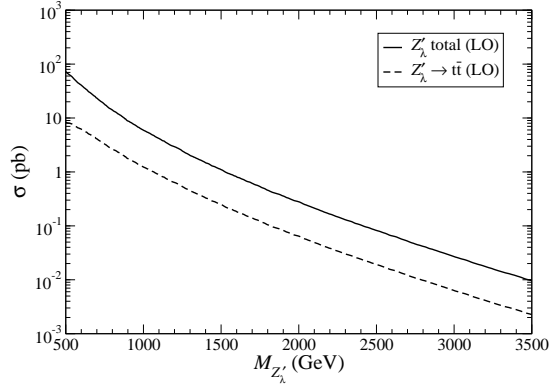}
\caption{Cross sections for $Z'_\lambda$ production (solid) and for $Z'_\lambda \to t \bar t$ (dashed) at LHC as a function of the $Z'_\lambda$ mass.}
\label{fig:cross-lhc}
\end{center}
\end{figure}
The total leading order (LO) $\zp$ production cross section at 
LHC is plotted in Fig.~\ref{fig:cross-lhc} as a function of $M_\zp$, 
together with the maximum ({\em i.e.} when $\zp$ decays
only to SM fermions) cross section for $t \bar t$ final states.
The coupling constant of the new
$\mathrm{U}(1)'$ has been fixed for reference as
$g' = \sqrt{5/3}\, g_Y = \sqrt{5/3}\, g \,s_W/c_W$,
and cross sections are calculated using CTEQ5L parton distribution functions
\cite{Lai:1999wy}.


For the simulations in the next section we also take a $\zp$ mass of 1 TeV,
above the Tevatron exclusion limits for this model~\cite{ttD0}. The signal is generated with Monte Carlo statistics of 300 fb$^{-1}$ and rescaled to 30 fb$^{-1}$.
SM backgrounds include $t \bar tnj$ (with $nj$ standing for $n$ jets at the parton level), single top, $W/Z nj$,
$W/Z t \bar t nj$, $W/Z b \bar b nj$, $W/Z c \bar c nj$, diboson and triboson production.
They are generated with a Monte Carlo statistics of 30 fb$^{-1}$ using \ALPGEN, taking 
$m_t = 175$ GeV, $M_H = 115$ GeV. (The complete list of processes and numbers of events 
generated can be found in Ref.~\cite{AguilarSaavedra:2009es}.)
Events are interfaced to {\tt Pythia} 6.4~\cite{Sjostrand:2006za} to add soft ISR and 
FSR and pile-up, and perform hadronisation. The MLM prescription is also used to 
perform the matching for the backgrounds, with default values for the matching 
parameters. A detailed investigation of the uncertainties related to the matching 
procedure with \ALPGENs for the main background processes has been presented in 
Refs.~\cite{Mangano:2006rw,Alwall:2007fs}. 
In order to simulate a real detector environment
we use the fast simulation {\tt AcerDET}~\cite{RichterWas:2002ch} which is a generic LHC detector simulation, neither of ATLAS nor of CMS, with standard settings.
In particular, the lepton isolation criteria require a separation $\Delta R > 0.4$ from other clusters and a maximum energy deposition $\Sigma E_T = 10$ GeV in a cone of $\Delta R = 0.2$ around the reconstructed electron or muon. Jets are reconstructed using a cone algorithm with $\Delta R = 0.4$. In this analysis we only focus on central jets with pseudo-rapidity $|\eta| < 2.5$.
For central jets, a simple $b$ tagging is performed with probabilities of 60\% for $b$ jets, 10\% for charm and 1\% for light jets. 

Our estimate of the signal relies on the LO approximation. In the absence of 
a proper next-to-leading order (NLO) calculation, we estimate the impact of 
hard NLO corrections using MCFM v5.6 NLO code~\cite{Campbell:2002tg}. 
We have rescaled $Z$ and $W$ masses (and accordingly the $G_F$ coupling constant) 
in such a way that the $Z$ mass is pushed to $500$~GeV and $1$~TeV, while 
keeping the SM couplings, in order to simulate the sequential $Z'_{SM}$. 
We selected the $b \bar b$ final state among the possible ones, choosing a window
$M_{Z'_{SM}} - 7 \Gamma_{Z'_{SM}} < m_{b \bar b} < M_{Z'_{SM}} + 7
\Gamma_{Z'_{SM}}$.
As a result the $K$-factor, defined as
$\sigma({\rm NLO}) / \sigma({\rm LO})$
with renormalization and factorization scales fixed to
$M_{Z'_{SM}}$, turns out to be of the order of
$20$\% for both $M_{Z'_{SM}} = 500$~GeV and $M_{Z'_{SM}} = 1$~TeV.

\section{\label{analysis} 
$Z' \to t \bar t$ and $Z' \to t \bar tj$ observation at LHC}

The presence of a $Z'$ resonance is detected as a bump in the $t \bar t$ invariant mass spectrum, which can eventually be normalised from real data with a good precision. For this analysis we restrict ourselves to the $t \bar t$ semileptonic decay channel, in which the kinematics can be fully reconstructed.
The pre-selection criteria are: (i) one charged lepton with $p_T > 30$ GeV; (ii) two $b$-tagged jets with $p_T > 20$ GeV; (iii) at least two light jets with $p_T > 20$ GeV; (iv) total transverse energy $H_T > 750$ GeV. The latter cut is implemented in order to reduce the QCD 
$t \bar t nj$ background, as well as to remove from the signal processes the SM component mediated by $Z$ and $\gamma ^*$. Note that these cuts can be optimised to enhance the statistical significance of the signal but, on the other hand, reducing the background too much makes its normalisation from real data difficult.
The number of events fulfilling these requirements are collected in Table~\ref{tab:nsnb-pre} (left) for the signal and main backgrounds (the rest of backgrounds are not explicitly shown but we keep them in the calculations). On the right panel we show the decomposition of the $t \bar t nj$ background in $n=0,\dots,5$ multiplicities. The dominance of the $n=1,2,3$ subsamples results from the $H_T$ cut, which has a much higher rejection for lower jet multiplicities.

\begin{table}[htb]
\begin{center}

\begin{tabular}{ccc}
\begin{tabular}{ccccc}
                   & Pre.  & \quad &                    & Pre.   \\[1mm]
$\Zlp$ ($0j$)      & 493.2 &       & $\Zlp$ ($1j$)      & 1213.9 \\
\hline
$t \bar tnj$       & 64688 &       & $t \bar tb \bar b$ & 657  \\
$tW$               & 2425  &       & $Wnj$              & 915    \\
$tj$               & 445   &       & $Wb \bar bnj$      & 2202
\end{tabular}
& \quad \quad &
\begin{tabular}{ccc}
             & Pre.  & acc. $H_T$ \\[1mm]
$t \bar t0j$ & 7512  & 2.7\% \\
$t \bar t1j$ & 16665 & 9.1\% \\
$t \bar t2j$ & 16392 & 22.7\% \\
$t \bar t3j$ & 10299 & 43.6\% \\
$t \bar t4j$ & 4580  & 66.6\%  \\
$t \bar t5j$ & 1601  & 84.6\%
\end{tabular}
\end{tabular}
\end{center}
\caption{Left: number of signal and background events at the pre-selection level. Right: 
decomposition of the $t \bar t nj$ background in subsamples with $n$ jets at the partonic level and acceptance of the $H_T$ cut. The luminosity is 30 fb$^{-1}$.}
\label{tab:nsnb-pre}
\end{table}

The first step to search for the $Z'$ signal is the reconstruction of the top quark pair.
These are found by choosing the best pairing between $b$ jets and reconstructed $W$ bosons:
\begin{enumerate}
\item The hadronic $W$ is obtained with the two jets (among the three ones with largest $p_T$) having an invariant mass closest to $M_W$.
\item The leptonic $W$ is obtained from the charged lepton and the missing energy, identifying $(p_\nu)_T = \ptmiss$, requiring $(p_{\ell}+p_\nu)^2 = M_W^2$ and solving for the longitudinal component of the neutrino momentum. If no real solution exists, the neutrino transverse momentum is decreased in steps of 1\% and the procedure is repeated. If no solution is still found after 100 iterations, the discriminant of the quadratic equation is set to zero.
Both solutions for the neutrino momentum are kept, and the one giving best reconstructed masses is selected.
\item The two top quarks are each reconstructed with one of the $W$ bosons and one of the $b$ jets, and are labelled as `hadronic' and `leptonic' corresponding to the hadronic and leptonic $W$, respectively.
\item The combination minimising
\begin{small}
\begin{equation}
\frac{(m_W^\text{had}-M_W)^2}{\sigma_W^2} + 
\frac{(m_W^\text{lep}-M_W)^2}{\sigma_W^2} + 
\frac{(m_t^\text{had}-m_t)^2}{\sigma_t^2}
\frac{(m_t^\text{lep}-m_t)^2}{\sigma_t^2}
\end{equation}
\end{small}%
is selected, with $\sigma_W = 10$ GeV, $\sigma_t = 14$ GeV~\cite{Aad:2009wy}.
\end{enumerate}

\begin{figure}[t]
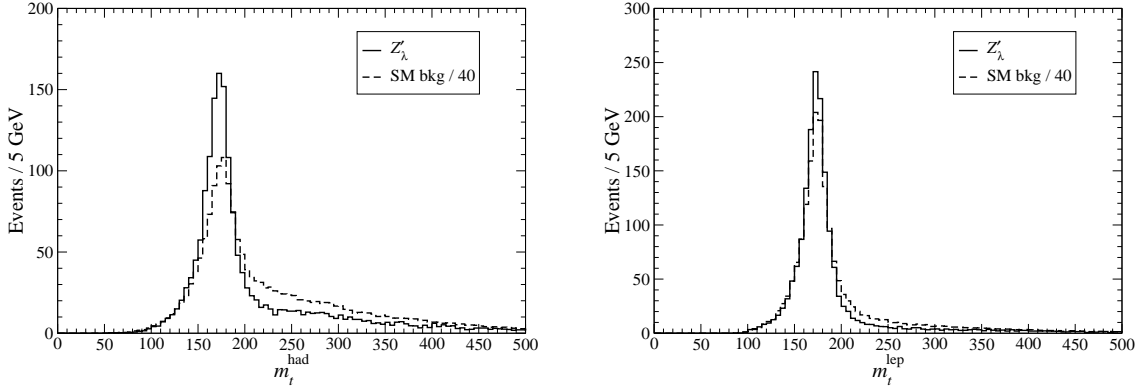

\begin{center}
\begin{tabular}{ccc}
\epsfig{file=Figs/mth-pre.eps,height=5.1cm,clip=} & \quad &
\epsfig{file=Figs/mtl-pre.eps,height=5.1cm,clip=}
\end{tabular}
\caption{Reconstructed top quark masses. 
The luminosity is 30 \fbin.}
\label{fig:mrec-pre}
\end{center}
\end{figure}

\begin{figure}[htb]
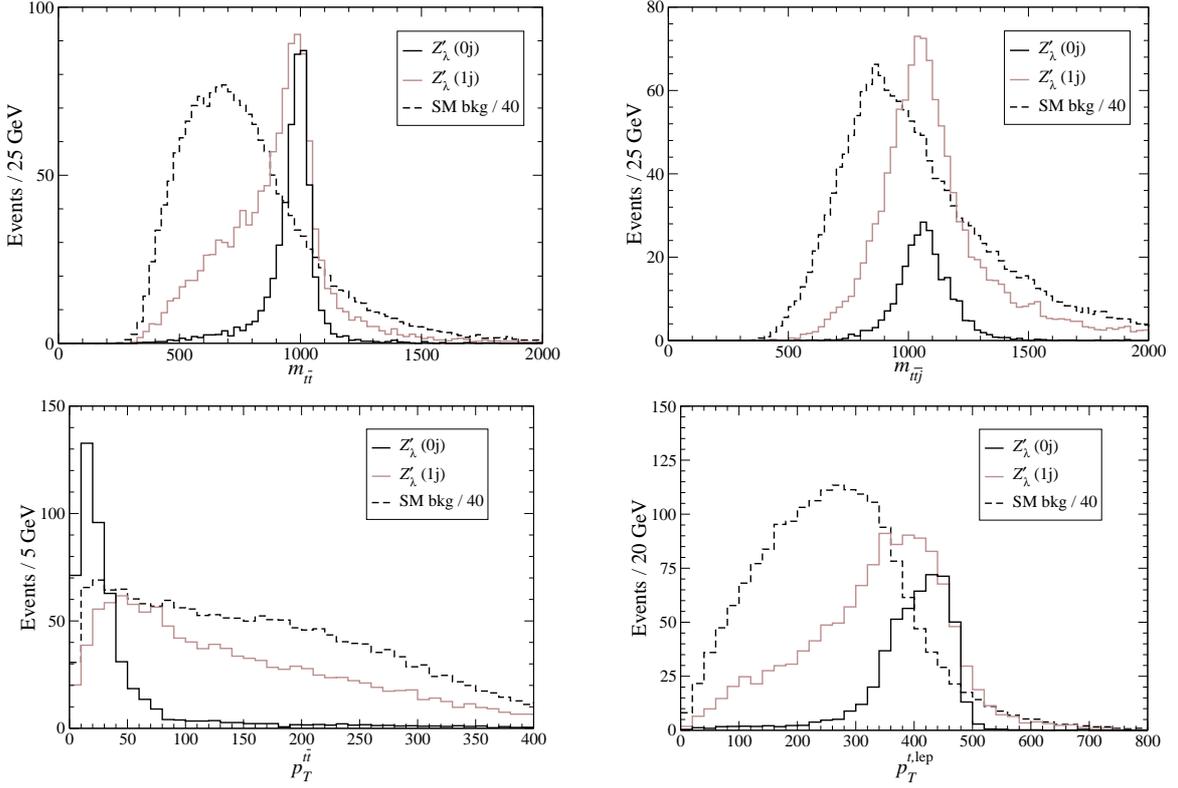

\begin{center}
\begin{tabular}{ccc}
\epsfig{file=Figs/mtt-pre.eps,height=5.1cm,clip=} & \quad &
\epsfig{file=Figs/mttj-pre.eps,height=5.1cm,clip=} \\
\epsfig{file=Figs/ptt-pre.eps,height=5.1cm,clip=} & &
\epsfig{file=Figs/ptl-pre.eps,height=5.1cm,clip=}
\end{tabular}
\caption{Kinematic distributions at pre-selection. Up, left: $t \bar t$ invariant mass; up, right: $t \bar t j$ invariant mass (requiring at least three light jets); down, left: transverse momentum of the $t \bar t$ pair; down, right: transverse momentum of the top quark decaying leptonically (not used for event selection). The luminosity is 30 \fbin.}
\label{fig:Zp-pre}
\end{center}
\end{figure}

\begin{figure}[htb]
\begin{center}
\epsfig{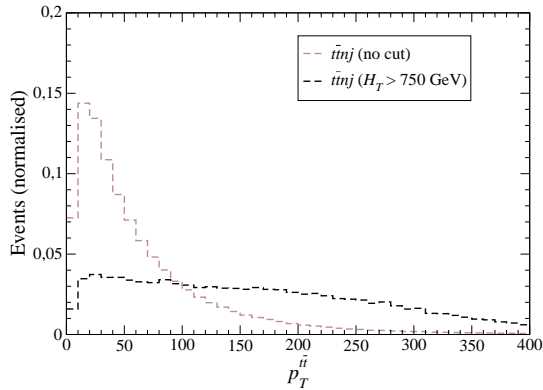}
\caption{$t \bar t$ transverse momentum distribution for the $t \bar t nj$ background at the pre-selection level with and without the $H_T$ cut. The luminosity is 30 \fbin.}
\label{fig:ptt-comp}
\end{center}
\end{figure}

We present the reconstructed mass distributions at the pre-selection level in Fig.~\ref{fig:mrec-pre}.
With the top quark pair identified, one can consider several variables to discriminate the signal from the background. The most interesting ones are the $t \bar t$, $t \bar t j$ invariant masses and the transverse momentum of the $t \bar t$ pair. They are presented in Fig.~\ref{fig:Zp-pre} for the two signal subsamples with zero ($0j$) and one ($1j$) jet at the partonic level and for the SM background. For completeness, we also show the transverse momentum distribution for the top quark decaying leptonically (for the hadronic top quark it is similar), which is not used to enhance the signal significance.\footnote{Although at pre-selection this variable seems to have a good discriminating power for $Z'$ ($0j$) events, after event selection criteria based on $p_T^{t \bar t}$ and other variables the signal and background distributions become very similar. Nevertheless, this and other distributions, as for example the $t \bar t$ pair rapidity, can be implemented in a likelihood function to achieve a better discriminating power than with the cut-based analysis implemented here. Such optimisation is beyond the scope of the present work.}
The most remarkable features, which guide our subsequent analysis, are:
\begin{enumerate}
\item The $m_{t \bar t}$ distribution has a large off-peak component from the $Z'$ ($1j$) sample. If the SM background cannot be predicted with very good accuracy (as it seems the case for $t \bar t$ plus several hard jets), and must be normalised from off-peak data, this tail will behave as a combinatorial background reducing the peak significance unless a specific analysis is carried out with a different reconstruction for these events.
\item The $Z'$ ($1j$) sample exhibits a good peak in the $t \bar t j$ invariant mass distribution, although slightly shifter towards $m_{t \bar t j}$ values larger than 1 TeV.
\item We also observe that 
the $t \bar t$ transverse momentum is typically much smaller for the $Z'$ ($0j$) signal than for the background. Although $t \bar t$ pairs are typically produced with low transverse momentum, the requirement $H_T > 750$ GeV has a higher suppression for lower hard jet multiplicities, as seen in Table~\ref{tab:nsnb-pre}.
As a result of this cut, the $\pttt$ distribution is shifted towards large values, as it can observed in Fig.~\ref{fig:ptt-comp}. This higher jet multiplicity is also the reason for the worse reconstruction of the hadronic top in Fig.~\ref{fig:mrec-pre}.
\end{enumerate}
The best strategy to maximise the signal significance is to develop two analyses. The first one aims to reconstruct the $t \bar t$ peak in the region of low transverse momenta of the $t \bar t$ pair setting $\pttt < 50$ GeV, which eliminates a large fraction of background and the off-peak signal contribution which would otherwise constitute a combinatorial background. The second analysis will search for the $t \bar tj$ peak in the complementary region $\pttt > 50$ GeV. We present these two analyses in turn. Then, for comparison, we will perform an inclusive analysis without separate reconstructions.

\subsection{Analysis I}

\begin{figure}[tb]
\begin{center}
\epsfig{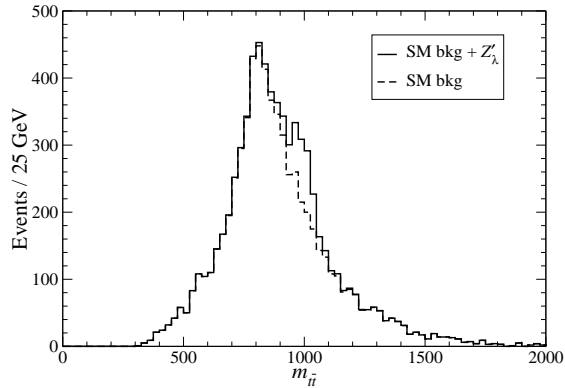}
\caption{$t \bar t$ invariant mass for the SM background and the background plus the $Z'$ signal at the selection level (analysis I). The luminosity is 30 \fbin.}
\label{fig:mtt-anI}
\end{center}
\end{figure}

As selection criteria we impose some loose quality cuts on reconstructed top quark masses and, more importantly, small transverse momentum for the $t \bar t$ pair,
\begin{eqnarray}
& 125~\text{GeV} < m_t^\text{had},m_t^\text{lep} < 225~\text{GeV} \,, \notag \\
& \pttt < 50~\text{GeV} \,.
\end{eqnarray}
The number of signal and background events with these cuts can be read in Table~\ref{tab:nsnb-anI}. 
\begin{table}[htb]
\begin{center}
\begin{tabular}{ccccccc}
                   & Sel.  & Peak  & \quad &                    & Sel.  & Peak \\[1mm]
$\Zlp$ ($0j$)      & 334.5 & 299.2 &       & $\Zlp$ ($1j$)      & 149.0 & 114.4 \\
\hline
$t \bar tnj$       & 6581  & 1652  &       & $t \bar tb \bar b$ & 15    & 3 \\
$tW$               & 85    & 17    &       & $Wnj$              & 39    & 13 \\
$tj$               & 9     & 4     &       & $Wb \bar bnj$      & 15    & 4 \\
\end{tabular}
\end{center}
\caption{Number of signal and background events at the selection level (analysis I) and at the $t \bar t$ resonance peak. The luminosity is 30 fb$^{-1}$.}
\label{tab:nsnb-anI}
\end{table}
We observe that the $\pttt$ cut significantly reduces the $t \bar t nj$ background and removes 90\% of the $Z'$ signal with $n=1$, in particular the off-peak contribution. The presence of the $Z'$ resonance can be spotted with the analysis of the $t \bar t$ invariant mass distribution, presented in Fig.~\ref{fig:mtt-anI} for the background alone and the background plus the signal at the selection level.
The number of events at the peak
\begin{align}
900~\text{GeV} < m_{t \bar t} < 1100~\text{GeV}
\label{ec:peak-tt}
\end{align}
can be read in Table~\ref{tab:nsnb-anI}. We will assume that we normalise the background by off peak measurements, obtaining a scaling factor $\kappa = 1.05$ determined from the comparison of the distributions for signal plus background and background alone. 
Then, the statistical significance of the peak is $S'/\sqrt{B'}$, where
\begin{equation}
B' = \kappa B \quad \Rightarrow \quad S'=(S+B)-B'
\end{equation}
and $S$, $B$ are the true numbers of signal and background events. For the peak region in Eq.~(\ref{ec:peak-tt}), the excess of events amounts to $7.78\sigma$.

\subsection{Analysis II}

\begin{figure}[tb]
\begin{center}
\epsfig{file=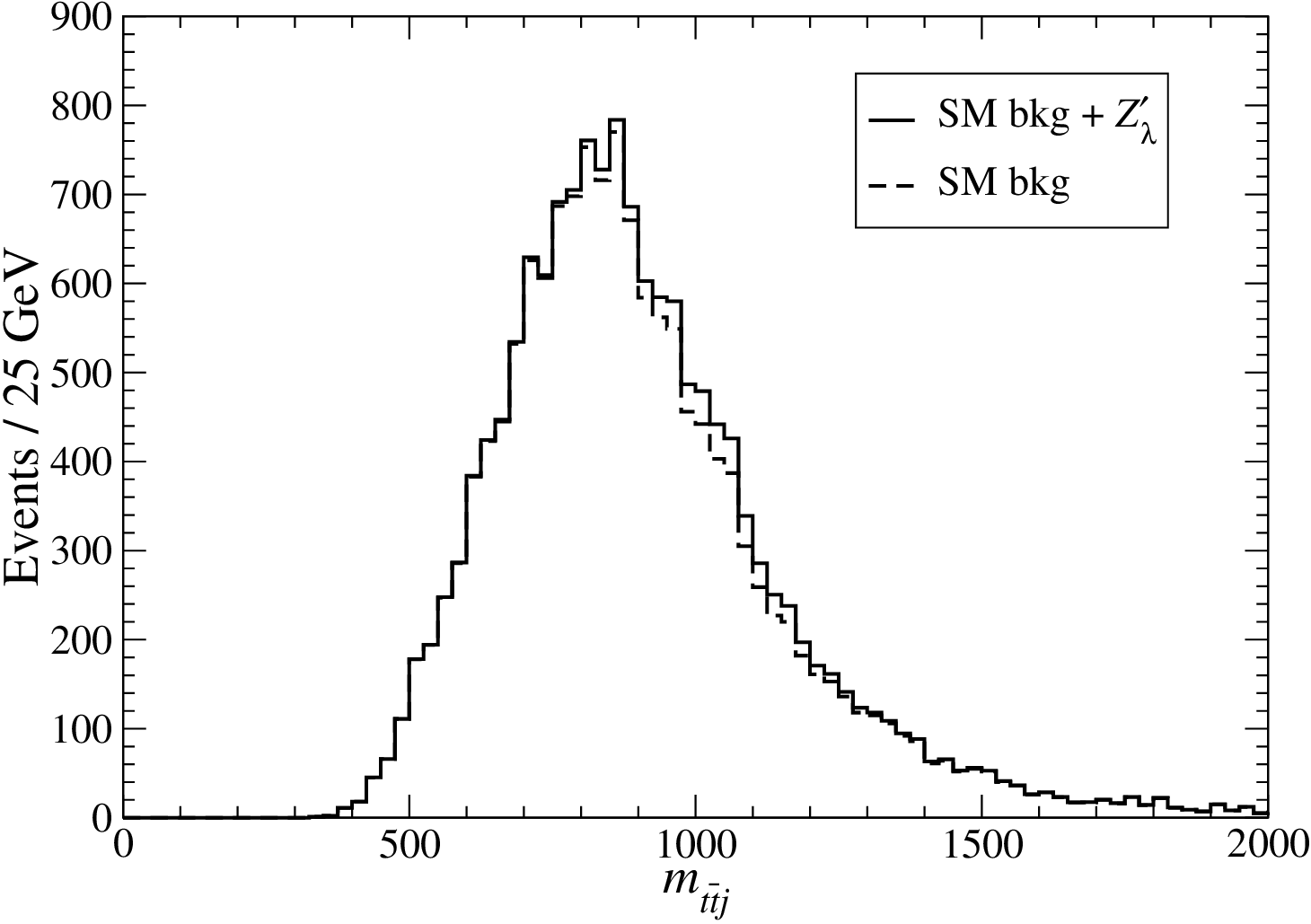,height=5.1cm,clip=}
\caption{$t \bar tj$ invariant mass for the SM background and the background plus the $Z'$ signal at the selection level (analysis II). The luminosity is 30 \fbin.}
\label{fig:mttj-anII}
\end{center}
\end{figure}

The main motivation for this analysis is the fact that
many of the $Z'$ ($1j$) events do not exhibit a peak in the $t \bar t$ invariant mass distribution and would fall off the peak region in Eq.~(\ref{ec:peak-tt}). This is seen clearly in Fig.~\ref{fig:Zp-pre} (up, left). When the invariant mass of the $t \bar t$ pair plus the hardest additional jet is considered (requiring in this case a minimum of three jets) the distribution exhibits a clear peak, although slightly displaced, as shown
on the upper right panel of that figure.
We then concentrate on the complementary event sample for this analysis, requiring at least three light jets and setting the cuts
\begin{eqnarray}
& 125~\text{GeV} < m_t^\text{had},m_t^\text{lep} < 225~\text{GeV} \,, \notag \\
& \pttt > 50~\text{GeV} \,, \notag \\
& \text{min}(\Delta R_{j,t_\text{had}},\Delta R_{j,t_\text{lep}}) < 1.6 
\end{eqnarray}
as selection criteria. The last one is implemented to reduce the background, since the signal events peaking at $m_{t \bar t j} \sim M_\zp$ are produced by FSR in $Z' \to t \bar t$. The number of events after these cuts are given in Table~\ref{tab:nsnb-anII}.
\begin{table}[htb]
\begin{center}
\begin{tabular}{ccccccc}
                   & Sel.  & Peak  & \quad &                    & Sel.  & Peak \\[1mm]
$\Zlp$ ($0j$)      & 52.2  & 41.7  &       & $\Zlp$ ($1j$)      & 412.8 & 293.5 \\
\hline
$t \bar tnj$       & 14226 & 4432  &       & $t \bar tb \bar b$ & 77    & 18 \\
$tW$               & 229   & 60    &       & $Wnj$              & 43    & 17 \\
$tj$               & 20    & 9     &       & $Wb \bar bnj$      & 77    & 25 \\
\end{tabular}
\end{center}
\caption{Number of signal and background events at the selection level (analysis II) and at the $t \bar tj$ resonance peak. The luminosity is 30 fb$^{-1}$.}
\label{tab:nsnb-anII}
\end{table}
The $t \bar t j$ distribution for the signal plus background and background alone is presented in Fig.~\ref{fig:mttj-anII}.
The number of events at the peak
\begin{align}
900~\text{GeV} < m_{t \bar tj} < 1200~\text{GeV}
\label{ec:peak-ttj}
\end{align}
can be read in Table~\ref{tab:nsnb-anII}. The background scaling factor in this case is $\kappa = 1.022$, and the excess of events over the (normalised) SM background expectation amounts to $3.56\sigma$.

\subsection{Inclusive analysis}

In order to see the advantage of separate, dedicated analyses for $Z' \to t \bar t$, $Z' \to t \bar tj$, we also perform the inclusive analysis for the two $Z'$ signal components, searching for a peak in the $m_{t \bar t}$ distribution. For event selection we only require a good reconstruction of the top quark pair,
\begin{align}
& 125~\text{GeV} < m_t^\text{had},m_t^\text{lep} < 225~\text{GeV} \,,
\end{align}
and drop the other kinematic cuts.
The number of signal and background events can be read in Table~\ref{tab:nsnb-an0}. 

\begin{table}[htb]
\begin{center}
\begin{tabular}{ccccccc}
                   & Sel.  & Peak  & \quad &                    & Sel.  & Peak \\[1mm]
$\Zlp$ ($0j$)      & 400.1 & 332.6 &       & $\Zlp$ ($1j$)      & 727.1 & 361.0 \\
\hline
$t \bar tnj$       & 35203 & 4196  &       & $t \bar tb \bar b$ & 163   & 9 \\
$tW$               & 567   & 40    &       & $Wnj$              & 170   & 20 \\
$tj$               & 109   & 12    &       & $Wb \bar bnj$      & 263   & 13 \\
\end{tabular}
\end{center}
\caption{Number of signal and background events at the selection level (inclusive analysis) and at the $t \bar t$ resonance peak. The luminosity is 30 fb$^{-1}$.}
\label{tab:nsnb-an0}
\end{table}

\begin{figure}[htb]
\begin{center}
\epsfig{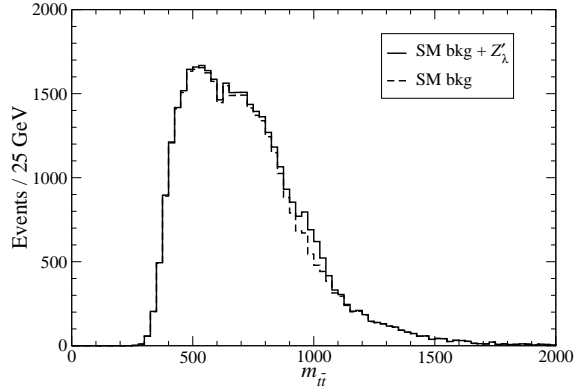}
\caption{$t \bar t$ invariant mass for the SM background and the background plus the $Z'$ signal at the selection level (inclusive analysis). The luminosity is 30 \fbin.}
\label{fig:mtt-an0}
\end{center}
\end{figure}

The $t \bar t$ distribution for the signal plus background and background alone is presented in Fig.~\ref{fig:mtt-an0}.
The number of events at the peak
\begin{align}
900~\text{GeV} < m_{t \bar t} < 1100~\text{GeV}
\label{ec:peak-tt-2}
\end{align}
is also given in Table~\ref{tab:nsnb-an0}. As expected, in this case the background scaling factor is larger than for the other analyses, $\kappa = 1.055$, and the peak significance is of $6.77\sigma$. This number is 25\% smaller than the combined significances of the other two analyses, $8.55\sigma$.

\section{Conclusions}

We provide a new code implemented in ALPGEN for neutral vector resonance production, 
including the higher order corrections from real emission and the virtual LL contributions.
In order to illustrate its features we have generated $t \bar t$ and $t \bar tj$ events by the  
exchange of a leptophobic gauge boson based on $\text{E}_6$ at LHC.
The analysis of the $Z'_\lambda \to t \bar t$ and $Z'_\lambda \to t \bar t j$  
samples show that such a program is necessary to account for $\sim 23$\% of 
the $t \bar t X$ events. Indeed, for these events the resonance is not found in the $t \bar t$ invariant mass distribution but on the $t \bar t j$ invariant mass. Therefore, their presence cannot be accounted for by a $K$ factor multiplying the LO $\zp \to t \bar t$ cross section but has to be properly simulated at the generator level.

We have shown that the LHC discovery potential for neutral vector resonances in the $t \bar tX$ final state benefits from a separate analysis for $t \bar t$ and $t \bar tj$. For the case examined (with a $\zp$ mass of 1 TeV), the enhancement over an inclusive search for $t \bar t$ resonances is of a 25\% in the statistical significance. This improvement is expected to be maintained for larger masses, and of course for other $Z'$ models than the leptophobic $\zp$ one used in the simulations. For other $t \bar t$ resonances the trend is expected to be the same as well.

Finally, the inclusion of corrections to $Z' \to t \bar t$ not only translates into a better discovery potential but also into a proper description of the kinematical distributions.
A code implementing them is necessary to establish the model
once a new vector resonance is discovered, discriminating
among models by a more precise measurement of the angular
distributions~\cite{Bernreuther:1997gs} and the different $Z'$
couplings~\cite{Diener:2009vq,DelAguila:1995fa}.

\section*{Acknowledgements}
This work has been partially supported by 
MICINN 
(FPA2006-05294), by Junta de Andaluc\'{\i}a (FQM 101,  FQM 437 and FQM 03048),
and by the European Community's Marie-Curie Research Training
Network under contract MRTN-CT-2006-035505 ``Tools and Precision
Calculations for Physics Discoveries at Colliders''.
The work of J.A.A.S. and M.T. has been supported by MICINN Ram\'on y Cajal and Juan de la Cierva contracts, respectively.
F.P. and M.M. acknowledge 
warm hospitality and partial support from CERN, Physics Department, TH Unit, 
during the completion 
of this work. 

\end{document}